\documentclass[prb,aps,twocolumn,nopacs,superscriptaddress,nofootinbib,reprint]{revtex4}
\usepackage{graphicx}
\usepackage{dcolumn}
\usepackage{bm}
\usepackage{amsmath}
\usepackage{epsfig}
\usepackage{bbm}
\usepackage{mathrsfs}
\usepackage{amsfonts}
\usepackage{mathtools}
\usepackage{color} 
\usepackage[colorlinks=true,linkcolor=blue,urlcolor=cyan,citecolor=blue]{hyperref}
\usepackage{multirow}  
\usepackage{braket}

\def\ii{{\rm i}}  \def\ee{{\rm e}}
  
\newcommand{\abs}[1] {\mathopen{}\left|#1\right|\mathclose{}}

\newcommand{\ccpar}[1] {\mathopen{}\left(#1\right)\mathclose{}}

    \def\ab{{\boldsymbol{a}}}

    \def\da{\ddot p}
  
      \def\Ef{{E_{\rm F}}}

\def\ww{{\omega}}  \def\vep{\varepsilon}  \def\sw{{s\omega}}

\begin{document}
\title{Strong-field-driven dynamics and high-harmonic generation in interacting 1D systems}

\author{Sandra~de~Vega}
\affiliation{ICFO-Institut de Ciencies Fotoniques, The Barcelona Institute of Science and Technology, 08860 Castelldefels (Barcelona), Spain}
\author{Joel~D.~Cox}
\email{cox@mci.sdu.dk}
\affiliation{Center for Nano Optics, University of Southern Denmark, Campusvej 55, DK-5230 Odense M, Denmark}
\affiliation{Danish Institute for Advanced Study, University of Southern Denmark, Campusvej 55, DK-5230 Odense M, Denmark}
\author{Fernando~Sols}
\affiliation{Departamento de F\'isica de Materiales, Facultad de Ciencias F\'isicas, Universidad Complutense de Madrid, E-28040 Madrid, Spain}
\affiliation{IMDEA (Instituto Madrile\~no de Estudios Avanzados) Nanociencia, Cantoblanco 28049, Madrid, Spain}
\author{F.~Javier~Garc\'{\i}a~de~Abajo}
\email{javier.garciadeabajo@nanophotonics.es}
\affiliation{ICFO-Institut de Ciencies Fotoniques,  The Barcelona Institute of Science and Technology, 08860 Castelldefels (Barcelona), Spain}
\affiliation{ICREA-Instituci\'o Catalana de Recerca i Estudis Avan\c{c}ats, Passeig Llu\'{\i}s Companys, 23, 08010 Barcelona, Spain}

\date{\today}

\begin{abstract}
The observation of high-order harmonic generation (HHG) from bulk crystals is stimulating substantial efforts to understand the involved mechanisms and their analogue to the intuitive three-step recollision model of gas phase HHG. On the technological side, efficient solid-state HHG is anticipated to enable compact attosecond and ultraviolet light sources that could unveil electron dynamics in chemical reactions and provide sharper tomographic imaging of molecular orbitals. Here we explore the roles of electronic band structure and Coulomb interactions in solid-state HHG by studying the optical response of linear atomic chains and carbon nanotubes to intense ultrashort pulses. Specifically, we simulate electron dynamics by solving the single-particle density matrix equation of motion in the presence of intense ultrafast optical fields, incorporating tight-binding electronic states and a self-consistent electron-electron interaction. While linear atomic chains constitute an idealized system, our realistic 1D model readily provides insight on the temporal evolution of electronic states in reciprocal space, both in the absence or presence of electron interactions, which we demonstrate to play an important role in the HHG yield. This model further predicts that doped semiconductors generate high harmonics more efficiently than their metallic and undoped counterparts. To complement this idealized system we also show results for HHG in more realistic quasi-1D structures such as carbon nanotubes, the behavior of which is found to be in good qualitative agreement with the atomic chains. Our findings apply directly to extreme nonlinear optical phenomena in atoms on surfaces, carbon-based structures, linear arrays of dopant atoms in semiconductors, and linear molecules, such as polycyclic aromatic hydrocarbon chains, and can be straightforwardly extended to optimize existing platforms for HHG or identify new solid-state alternatives in the context of nonlinear plasmonics.
\end{abstract}

\maketitle

\section{Introduction} \label{sec1}

High-harmonic generation (HHG) is perhaps the most striking example of a nonlinear optical process and its ability to spectrally and temporally disperse intense laser light.\cite{BK00,CK07} Initial reports of HHG from atomic gases\cite{BBR1977} revealed a light emission intensity plateau extending over many integer multiples of the fundamental exciting laser frequency and characterized by an abrupt drop at a specific cutoff energy. Concise theoretical explanations of the underlying physics were developed shortly thereafter, culminating in the celebrated three-step model of an atom interacting with a single cycle of an intense impinging optical field: tunnel ionization triggered by the driving electric field liberates an electron from the atom that gains additional kinetic energy as it is driven away from and back towards the parent nucleus, ultimately emitting, upon recollision (through coherent interactions of the electron wave function with itself), light at high harmonic orders of the fundamental frequency.\cite{MGJ1987,FLL1988,LBI94} This extreme nonlinear optical phenomenon is a source of coherent high-frequency electromagnetic radiation, which can be processed to produce attosecond optical pulses, thus garnering significant attention as the means to develop micron-scale XUV-light sources\cite{MK10} (i.e., the equivalent of table-top synchrotrons) and perform quantum logic operations at optical clock rates,\cite{HLO15} while enabling visualization of electronic band structures,\cite{VHT15,WHC16} monitoring electron-hole recollisions in real time,\cite{ZLS12} resolving subfemtosecond processes governing chemical reactions,\cite{WPL16} and recording electron dynamics in molecular orbitals.\cite{Z00,FHC19}

Despite the numerous fascinating advances in science and technology that have resulted from atomic HHG, the expense and delicacy of the associated experimental set-ups renders their use hardly practical outside of specialized laboratory facilities. In contrast, recent observations of HHG from solids\cite{GDS11,SHL14,LGK15} are establishing new paths for attosecond science and strong-field physics, potentially leading towards XUV and attosecond light sources in compact solid-state devices. While the three-step recollision model\cite{C93_2} offers an intuitive understanding of HHG from atoms in the gas phase, the picture is less clear for solid-state HHG. Early theoretical proposals considered a three-step-like model in which an electron undergoes Bloch oscillations within an electronic band (either valence or conduction after interband tunneling) as a consequence of the change in direction of acceleration after half an optical cycle of the driving electric field;\cite{HKK17,OCO17,GR19} subsequently the excited electron scatters within its band (i.e., intraband HHG) or recombines with the parent hole or ion (i.e., interband HHG), and finally, it recollides with the first- and second-nearest holes or ions. 
However, this simplified description does not explain the role of electron-electron correlations, and furthermore, available experiments and numerical simulations often do not elucidate the specific origin of generated harmonics (e.g., from interband or intraband charge-carrier motion); the generation of even-order harmonics, the existence of atto-chirps, the formation of a well-defined high-energy cutoff, and numerous aspects of the electronic band structure still remain underexplored in the context of solid-state HHG.\cite{KMF13,VMO14,LW16,MVO15}
 
Further insight into the aforementioned open questions in HHG from condensed-matter systems can be gathered by analyzing one of the simplest models in solid-state physics: the Su-Schrieffer-Heeger (SSH) chain,\cite{SSH1979,S1985_2,GBS97} consisting of a dimerized linear chain of atoms described in the tight-binding approximation, with alternating hopping energies assigned to each of the two neighboring atoms on side of any given atom [Fig.\,\ref{Fig1}(a)]. As we discuss below, the SSH model is a convenient system to explore electronic band structure effects in the optical response of materials, as appropriate choices of hopping energies reveal either metallic, insulating, or topologically insulating behavior. To explore the effect of topology on HHG, recent works\cite{BH18,JB19} have employed the SSH tight-binding model and its analogue in more rigorous time-dependent density functional theory (TDDFT) simulations of atomic chains, predicting improved harmonic yields associated with the topolotical insulator (TI) phases for sub-band-gap photon energies that are robust under distortions, continuous phase transitions, and choice of on-site potentials.\cite{DB19} In a related study,\cite{HBM18} the transition from atomic-like systems to solid-state bulk materials was analyzed in the context of HHG, emphasizing the evolution in cutoff energy as the chain length increases, and concluding that a chain of six atoms constitutes the optimal length for this transition to occur as a consequence of changes in the state density.

Seeking to optimize HHG yields in condensed-matter systems, we explore the synergy between electronic band structure and optical resonances in finite SSH chains, which constitute a convenient, computationally inexpensive model that has already been demonstrated to qualitatively describe HHG predicted in the more rigorous TDDFT simulations of related 1D systems.\cite{BH18,JB19} We augment the SSH tight-binding Hamiltonian with a term accounting for electron-electron interactions, incorporating a single-electron density matrix description of the optical response and introducing the effect of inelastic charge-carrier scattering through a phenomenological damping rate; this prescription allows us to systematically explore the dependence of HHG yield on the spectral characteristics of the impinging optical pulse and identify frequencies at which HHG is enhanced by optical resonances associated with the electronic band gap or collective electron motion (i.e., plasmons) in SSH chains. We further explore the effect of electrical doping on HHG by populating the electronic bands with additional charge carriers; the added charges can Pauli-block specific electronic transitions and introduce collective resonances, thus facilitating explorations of both. In order to verify the qualitative predictions based on the SSH model in a more realistic condensed-matter platform, we investigate HHG in finite carbon nanotubes (CNTs) of various chiralities that produce similar electronic band features and also display different electronic behavior (metallic, insulating, and topologically insulating). Our findings elucidate the roles of these features intrinsic to different solid-state systems, providing a road map for the identification and engineering of next-generation solid-state nonlinear optical devices, with a view to producing XUV light and/or attosecond pulses.

\section{Electron dynamics}

In our SSH model, spin-degenerate electrons occupy the orbitals $\ket{l}$ located at atomic sites $x_l=la$ uniformly spaced with the lattice constant $a$. Single-electron states $\ket{\varphi_j}$ with associated energy eigenvalues $\hbar\vep_j$ satisfying $\hbar\vep_j\ket{\varphi_j}=H_0\ket{\varphi_j}$ are then obtained by expanding in the site basis according to $\ket{\varphi_j}=\sum_l a_{jl}\ket{l}$, where $a_{jl}$ are real-valued expansion coefficients. Following the formalism introduced elsewhere to simulate the optical response of graphene nanoislands,\cite{paper183,paper247} the electron dynamics is described by the single-particle density matrix $\rho = \sum_{ll'} \rho_{ll'} \ket{l}\bra{l'}$ constructed from time-dependent matrix elements $\rho_{ll'}$ and governed by the equation of motion
\begin{equation} \label{eq:eom}
	\dot\rho = - \dfrac{\ii}{\hbar} \left[ H_0 - e\phi, \rho \right] -  \frac{\gamma}{2} \left( \rho - \rho^0 \right),
\end{equation}
where $\rho^0$ denotes the equilibrium density matrix to which the system relaxes at a phenomenological rate $\hbar\gamma=50$\,meV (i.e., a relaxation time $\tau = \gamma^{-1} \sim 13.2$\,fs) and $\phi=\phi^{\rm ext} + \phi^{\rm ind}$ is the electrostatic potential, which includes contributions from both the impinging light electric field, $\phi_l^\text{ext}=- x_l E(t)$, and the electron-electron (e-e) interaction, $\phi_l^\text{ind} = \sum_{l'}v_{ll'} \rho^{\rm ind}_{l'}$; the latter quantity renders the equation of motion self-consistent through its dependence on the induced charge $\rho^{\rm ind}_l=-2e\ccpar{\rho_{ll}-\rho^0_{ll}}$ (the factor of 2 accounts for spin degeneracy) mediated by the spatial dependence of the Coulomb interaction $v_{ll'}$ between atoms $l$ and $l'$, for which we choose parameters associated with carbon 2p orbitals.\cite{paper183} In Fig.\,\ref{Fig1}(b) we plot the employed Coulomb interaction compared to that of a point-like charge. The equilibrium density matrix is constructed in the state representation according to $\rho^0_{jj'} = \delta_{jj'} f_j$, where $f_j$ is the occupation factor of state $\ket{\varphi_j}$ according to the Fermi-Dirac statistics (we assume zero temperature), and transformed to site representation through $\rho_{ll'}=\sum_{jj'} a_{jl} a_{j'l'}\rho_{jj'}$.

Incidentally, linear response theory [obtained by replacing the $[\phi,\rho]$ term by $[\phi,\rho^0]$ in Eq.\ (\ref{eq:eom})] yields a solution to Eq.\ (\ref{eq:eom}) for a monochromatic external electric field $E^{\rm ext}\ee^{-\ii\ww t}+{\rm c.c.}$ of frequency $\omega$ in the form of the harmonic density matrix component $\rho^{(1)} \,\ee^{-\ii\omega t}$; the induced charge density $\rho^{\rm ind} = -2e \rho^{(1)}_{ll'}$ (with a factor of 2 for spin degeneracy) is then self-consistently computed in the so-called random-phase approximation\cite{PB1952,HL1970} (RPA) as $\rho^{\rm ind} = \chi^{(0)} \phi$, where
\begin{equation}\label{eq:chi0}
	\chi^{(0)}_{ll'} = \dfrac{2e}{\hbar}\sum_{jj'}\ccpar{f_{j'} - f_j} \dfrac{a_{jl}a_{j'l}a_{jl'}a_{j'l'}}{\omega + \ii\gamma/2 - \ccpar{\vep_j - \vep_{j'}}}
\end{equation}
is the non-interacting RPA susceptibility. The poles of $\chi^{(0)}$ are related to individual electron-hole (e-h) pair excitations, so that omission of the induced charge by taking $\phi\to\phi^{\rm ext}$ yields a response comprised of Lorentzian peaks at the energies $\hbar(\vep_j-\vep_{j'})$; including the self-consistent potential, we isolate the induced charge as $\rho^{\rm ind} = \chi \phi^{\rm ext}$, where the response function $\chi=\chi^{(0)}\left[ 1 - \chi^{(0)} v \right]^{-1}$ introduces new poles associated with collective charge carrier excitations through the Coulomb interaction. For simplicity, we neglect exchange interaction and spin effects.

Going beyond linear response, we solve the equation of motion through either of two complementary approaches that allow us to investigate the nonlinear optical response in different regimes. In the first method we resort to direct numerical integration of Eq.\ (\ref{eq:eom}) in the time domain (TD) to obtain the induced dipole moment
\begin{equation}
	p(t) = -2e\sum_l\rho^{\rm ind}_l x_l
\end{equation}
produced by various types of external fields $E(t)$ [e.g., continuous wave (cw) illumination or ultrashort pulses], from which Fourier transformation of $p(t)$ reveals its spectral decomposition and characterizes the optical response. The TD approach does not impose any limit on the strength or type of impinging field, thus enabling the study of the intensity-dependent optical response, including simultaneously the effects of saturable absorption and high-order harmonic generation. As we are primarily interested in the latter phenomenon, we quantify the radiation emitted from the SSH chain by the dipole acceleration, $\ddot p(\omega) = |\omega^2 p(\omega)|^2$.\cite{JB19,BM11,BH18,HBM18}

In the second approach, we assume monochromatic illumination (as in the RPA) and perturbatively expand the density matrix entering Eq.\ (\ref{eq:eom}) according to
\begin{equation}\nonumber
	\rho = \sum_{n=0}^\infty\sum_{s=-n}^n \rho^{n s} \ee^{\ii s \omega t},
\end{equation}
where $n=1, 2,\dots$ denotes the perturbation order and $s$ the harmonic index, such that $\rho^{ns}$ is defined only when $\abs{s}\leq n$. We then obtain the polarizabilities $\alpha^{(n)}_\sw$ as
\begin{equation}\label{eq:Pert-a-ns}
	\alpha^{(n)}_\sw = -\dfrac{2e}{\ccpar{E_0}^n} \sum_l \rho_{ll}^{n s}\, x_l,
\end{equation}
where $\rho^{ns}$ is computed following the prescription in Ref.\ \onlinecite{paper247} that constitutes an extension of the linear RPA to higher perturbation orders. We employ this method to calculate the nonlinear polarizabilities $\alpha^{(3)}_\omega $ (i.e., the leading nonlinear contribution to the response at the fundamental frequency, which is associated with the Kerr nonlinearity and two-photon absorption).

\section{Su-Schrieffer-Heeger model} 

\begin{figure*}
	\begin{center}
		\includegraphics[width=1\textwidth]{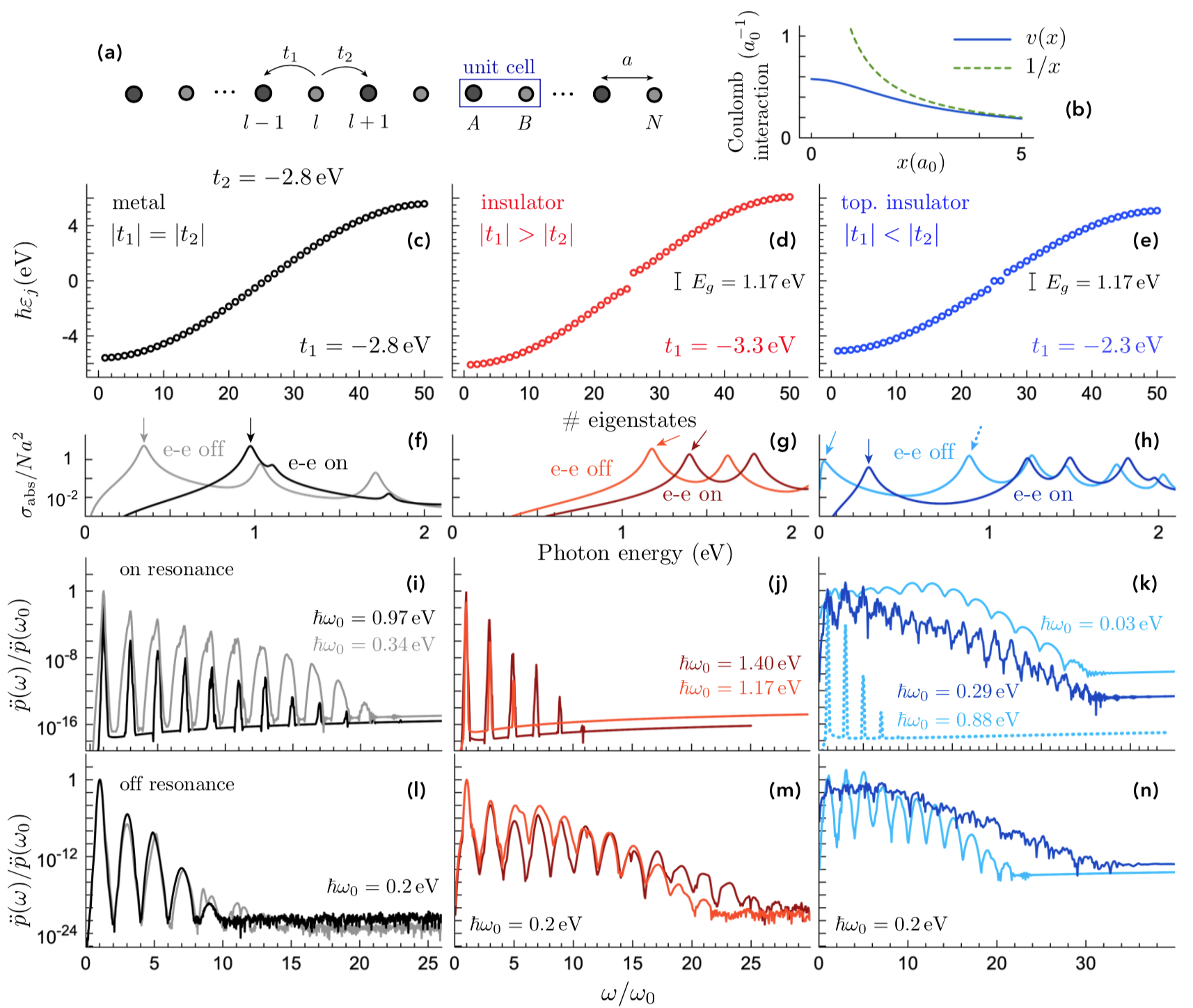}
		\caption{{\bf Characterizing the linear and nonlinear optical response of SSH chains.} {\bf (a)} Schematic representation of the SSH chain comprised of atoms A and B in the unit cell, with intracell (intercell) hopping $t_1$ ($t_2$), and uniform interatomic spacing $a$. {\bf (b)} Coulomb interaction of the SSH chain $v(x)$ compared to $1/x$. {\bf (c-e)} Band structure of SSH chains containing $N=50$ atoms, fixing the intercell hopping to $t_2=-2.8$\,eV and varying the intracell hopping $t_1$; depending on the choice of $t_1$ the system is (c) metallic ($t_1=t_2=-2.8$\,eV, black curves), (d) insulating ($t_1=-3.3$\,eV, red curves), and (e) a topological insulator (TI) ($t_1=-2.3$\,eV, blue curves). {\bf (f-h)} Normalized absorption cross section with electron-electron interactions switched on (e-e on) and off (e-e off) for the metal (f), the insulator (g), and the TI (h). {\bf (i-k)} Harmonic generation, quantified through the dipole acceleration $\ddot p(\omega) = |\omega^2 p(\omega)|^2$ (where $p(\omega)$ denotes the $\omega$ component of the induced dipole), produced by pulses of $10^{13}$\,W/m$^{2}$ peak intensity, 100\,fs FWHM duration, and carrier frequency $\omega_0$, with the latter quantities in each panel indicated by the color-coded legends and arrows in (f-h). {\bf (l-n)} Same as (i-k) but for a fixed pulse carrier energy of 0.2\,eV, away from the resonances appearing in (f-h).}
		\label{Fig1}
	\end{center}
\end{figure*}

Originally introduced to describe $p_z$ electrons in CH monomer chains (polyacetylene), the Su-Schrieffer-Heeger (SSH) model describes a 1D dimerized chain of $N$ atoms through a tight-binding (TB) Hamiltonian,\cite{SSH1979} and constitutes a simple yet powerful tool to explore non-trivial topological electronic band structure. We consider an SSH chain that contains $N/2$ unit cells with two sites per cell occupied by one atom from either sublattice $A$ or $B$ [Fig.\,\ref{Fig1}(a)]. Also, we denote the intracell and intercell hoppings as $t_1$ and $t_2$, respectively. The TB Hamiltonian describing the chain is\cite{AOP16}
\begin{align}\label{eq:H0_AandB}
	H_0 &= t_1 \sum_{l=1}^{N/2}\left( \ket{l,B} \bra{l,A } + {\rm h.c.} \right) \nonumber\\
	&+ t_2 \sum_{l=1}^{N/2-1}\left( \ket{l+1,A}\bra{l,B} + {\rm h.c.} \right),
\end{align}
which, expressed in a purely spatial representation, takes the form of a tridiagonal $N\times N$ matrix containing zeros along the diagonal and hoppings just above and below:
\begin{equation} \label{eq:H0}
	H_0=
	\begin{pmatrix}
		0    & t_1 & 0    & 0    & \dots  & 0 & 0 \\
		t_1  & 0   & t_2  & 0    & \dots  & 0 & 0 \\
		0    & t_2 & 0    & t_1  & \dots  & 0 & 0 \\
		0    & 0   & t_1  & 0    & \dots  & 0 & 0 \\
		\vdots     & \vdots      & \vdots & \ddots & \vdots & \vdots  \\
		0    & 0   & 0    & 0    & \dots  & 0 & t_1 \\
		0    & 0   & 0    & 0    & \dots  & t_1 & 0 \\
	\end{pmatrix}.
\end{equation}
The choice of hopping parameters determines the phase of the chain:\cite{AOP16} the band structure becomes metallic when $|t_1|=|t_2|$, insulating if $|t_1|>|t_2|$, and a TI (i.e., insulating in the bulk and with a edge states in the gap) when $|t_1|<|t_2|$\cite{AOP16}. Throughout this study we consider a chain with $N=50$ atoms located at the sites $ x_l = l a $ and having fixed intercell hopping $t_2=t_0$, choosing values $a=0.1421$\,nm and $t_0=-2.8$\,eV inspired by graphene. From the metallic chain ($t_1=t_2=t_0$), we perturb $t_1=t_0+0.5$\,eV to enter an insulating phase, whereas $t_1=t_0-0.5$\,eV yields the band structure of a TI. Diagonalization of $H_0$ reveals single-electron states characterized by the coefficients $a_{jl}$ (i.e., the amount of wavefunction $\ket{j}$ within the orbital at $x_l$) and energies $\hbar\vep_j$; we plot the electronic energies $\hbar\vep_j$ obtained for each of the three phases in Fig.\,\ref{Fig1}(c-e). With the chosen parameters, a band gap of energy $E_g = 1.17$\,eV emerges when $|t_1|\neq|t_2|$, with two quasi-degenerate states appearing in the middle of the band gap when $|t_1|<|t_2|$ [Fig.\,\ref{Fig1}(d)], corresponding to the edge states of the chain and giving the insulator its topological character.

\begin{figure*} 
	\begin{center}
		\includegraphics[width=1\textwidth]{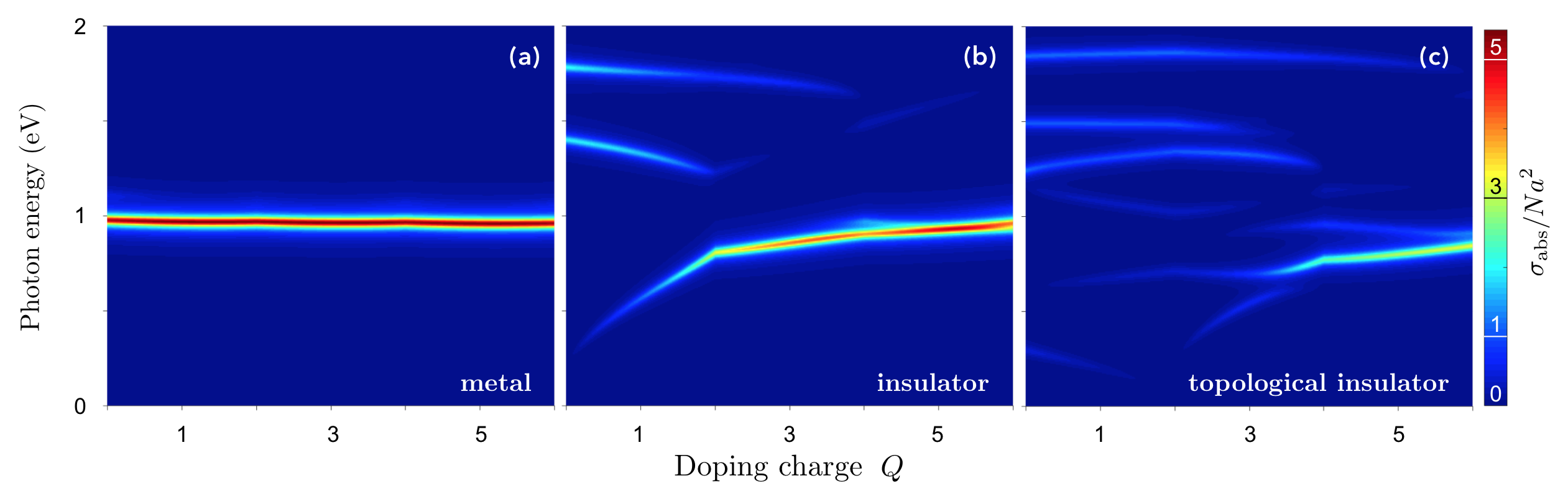}
		\caption{{\bf Effect of doping in the linear response of SSH chains.} Normalized absorption cross section as a function of frequency (vertical axis) and additional charge carriers $Q$ (horizontal axis) for the (a) metallic, (b) insulating, and (c) topologically insulating chains considered in Fig.\ \ref{Fig1}.
		}
		\label{Fig2}
	\end{center}
\end{figure*}

\section{Self-consistent interactions in the optical response of SSH chains}

Optical resonances, and plasmons in particular, are widely exploited in nano-optics to intensify local electromagnetic fields for a variety of applications, some of which involve the enhancement of nonlinear optical processes.\cite{KZ12,paper287} Here, we explore the ability of optical resonances in the three phases of the SSH model to drive HHG. In Fig.\ \ref{Fig1}(f-h) we identify through the linear absorption cross-section the available optical resonances in the metallic, insulating, and TI phases, both in situations when the self-consistent electron-electron (e-e) interaction is omitted and included. Neglecting e-e interactions, peaks in the absorption spectrum are associated with the energies of individual single-electron transitions (see discussion in Sec.\ \ref{sec1}), with amplitudes determined by their transition dipole moments. In contrast, if we include e-e interactions, the dominant transitions undergo large spectral blueshifts in all three phases. In Figs.\,\ref{Fig1}(f-h) we show the absorption cross section of the SSH chains, which we calculate through the optical theorem $\sigma_{\rm abs} = 4\pi(\omega/c) {\rm Im}\{\alpha^{(1)}_\omega\}$. We normalize the cross section to roughly the geometrical in-plane projection of the chain, $Na^2$. Incidentally, the edge states of the TI [Fig.\,\ref{Fig1}(h)] give rise to a low-energy resonance that does not appear for the insulator.

Given their importance in the linear response, it is expected that self-consistent e-e interactions also play a leading role in the nonlinear response. Figs.\ \ref{Fig1}(i-n) show normalized high harmonic spectra (quantified via dipole accelerations) produced by Gaussian pulses of  $10^{13}$\,W/m$^2$ peak intensity and 100\,fs full-width-at-half-maximum (FWHM) duration with central frequencies $\ww_0$ (i) targeting the dominant resonances in the linear spectra [Figs.\ \ref{Fig1}(i-k); see color-coded arrows indicating the energy of $\hbar\ww_0$] and (ii) off-resonance, with a frequency arbitrarily fixed to 0.2\,eV (i.e., away from optical resonances) in all cases [Figs.\ \ref{Fig1}(l-n)]. Resonant excitation of the metallic chain yields lower HHG when e-e interactions are included, presumably because charge screening in the metal damps the electron motion, while this effect is less important in the gapped systems. The number of observable harmonics is typically larger for lower-energy excitation and associated with more efficient interband generation, where the maximum cutoff energy in the non-interacting case is indicated by the largest available single-electron transition energy. Incidentally, the height of the first harmonic can vary widely from on-resonant to off-resonant conditions (e.g., by a factor up to $10^4$ in metallic chains).

Additionally, e-e interactions lead to collective optical resonances of higher strength compared with single-electron transitions, thus allowing us to reach HHG with significantly reduced intensity compared with previous studies that neglect those interactions.\cite{PBO94,OCO17}

Plasmons are associated with the motion of free electrons, and thus do not emerge in pristine semiconducting materials. However, in the 50-atom SSH chains that we consider here, the addition of only a few electron charges is sufficient to dramatically change the optical response; this phenomenon is explored in Fig.\ \ref{Fig2}, where we study the linear response in the RPA as a function of the doping charge $Q$ in all three SSH phases. Note that the charge is added in such a way that the free electrons equally populate available degenerate states. In contrast to the almost negligible electrical tuning for the metallic chain [Fig.\,\ref{Fig2}(a)], the insulating chains present overall a blueshift with increasing charge carrier density in the low-energy spectral features, which tend to coalesce into a prominent peak associated with intense optical absorption and a concentration of electromagnetic energy within the material.

In particular, at $Q=2$ in Fig.\,\ref{Fig2}(b) we observe a sharp feature that corresponds to the filling of the lowest unoccupied molecular orbital (LUMO); that is, the insulator gets free carriers in the conduction band and starts behaving as a metal (we note that a $Q=2$ doping corresponds to a Fermi level $\Ef\sim$0.81\,eV, which corresponds to the highest filled state in a one-electron picture, which is larger than the energy of the LUMO, $E_{\rm LUMO} = E_g/2 = 0.59$\,eV). For this reason, for $Q>2$ the main resonant feature begins to stabilize and by $Q=5$ ($\Ef\sim$1.11\,eV) it has coalesced in a prominent dipolar plasmon mode of frequency similar to that of the metallic chain because of the similar value of the density of states at the Fermi level in both cases. In contrast, for the TI chain [Fig.\,\ref{Fig1}(j)] we observe that quasi-degenerate edge states in the middle of the band gap produce a slight redshift of the main resonance and damp the strength of the absorption cross section, particularly at the LUMO energy ($Q=2$ or $\Ef\sim$0.63\,eV). At $Q=4$ the main resonance starts growing and by $Q=5$ the response is dominated by the plasmon, just like in the insulating chain.

\section{Intensity-dependent absorption}

The realization of HHG in solid-state systems necessitates optical pulses with peak intensities that cannot be sustained for long duration, lest the material be destroyed in the process. However, the interaction of extended pulses or cw fields with matter is appealing for technological applications relying on saturable absorption, an extreme nonlinear optical phenomenon that arises in all photonic materials and, like HHG, cannot be described in a perturbative framework.  

The enhanced light-matter interaction provided by optical resonances produces a more measurable absorption signal that facilitates detection of changes in the dielectric environment or the impinging light intensity; the latter effect is intensified by the concentration of electromagnetic energy in the material, which in turn can enhance its nonlinear optical response. Following this approach for HHG, we explore the nonlinear response associated with optical resonances of SSH chains by considering their interaction with intense impinging cw light, characterizing the optical response by the induced dipole moment at the fundamental frequency. Specifically, we extract the effective polarizability $\alpha_\ww$ by computing the induced dipole $p(t)$ in response to a monochromatic field $E(t)=E_0\ee^{-\ii\ww t}+{\rm c.c.}$ of intensity $I^{\rm ext} = (c/2\pi)|E_0|^2$. We then Fourier transform $p(t)$ over a single optical cycle to obtain
\begin{equation}
    \alpha_{\ww} = \frac{\omega}{2\pi E_0}\int^{t_{\rm cw}}_{{t_{\rm cw}}-2\pi/\omega}p(t)\ee^{-\ii\omega t}dt,
\end{equation}
where $t_{\rm cw}\gg \tau$ corresponds to a time at which the system has entered a steady state regime.

\begin{figure*} 
	\begin{center}
		\includegraphics[width=1\textwidth]{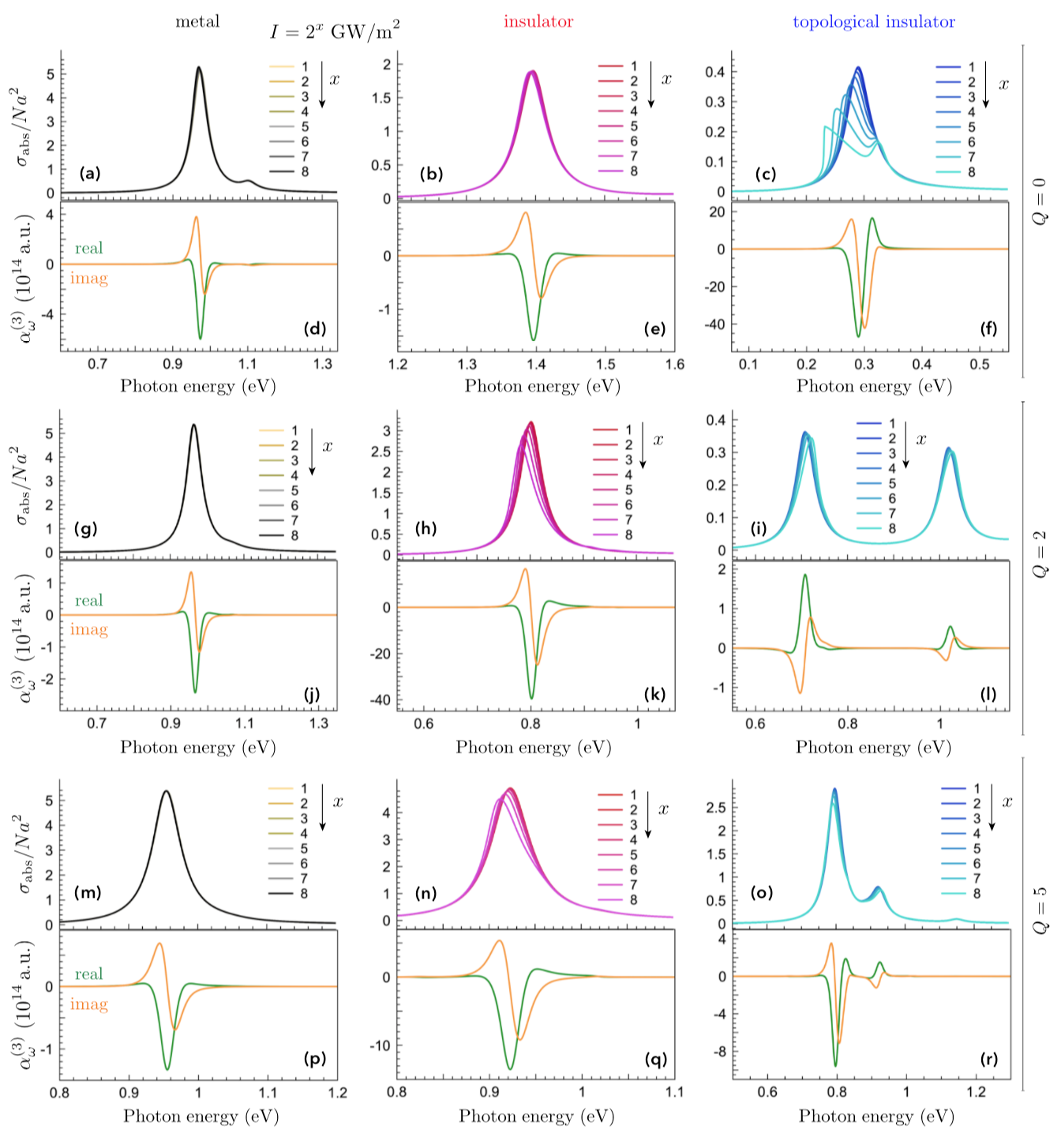}
		\caption{{\bf Intensity-dependent nonlinear absorption in SSH chains.} {\bf (a-c)} Normalized absorption cross-section $\sigma_{\rm abs}/Na^2$ simulated in the time domain for monochromatic cw illumination of increasing intensity near the optical resonances of undoped (a) metallic, (b) insulating, and (c) topologically insulating SSH chains. {\bf (d-f)} Perturbative solutions of the third-order polarizabilities associated with the optical Kerr nonlinearity $\alpha^{(3)}_\omega$, presented in atomic units for the corresponding SSH chains in (a-c), with real and imaginary parts indicated by green and orange curves, respectively. {\bf (g-l)} Same as (a-f) but for SSH chains doped with 2 additional electrons ($Q=2$). {\bf (m-r)} Same as (g-l) but for SSH chains doped with 5 additional electrons ($Q=5$).}
		\label{Fig3}
	\end{center}
\end{figure*}

In Fig.\,\ref{Fig3} we study the dependence on pulse intensity of the main resonances in the absorption spectra (upper rows) of the different SSH chains in relation to their corresponding Kerr polarizabilities $\alpha^{(3)}_\omega$, that is, with perturbation order $n=3$ and harmonic index $s=1$ (lower rows), for three different dopings: undoped [Figs.\ \ref{Fig3}(a-f)], LUMO doping (i.e., with two additional electrons, $Q=2$) [Figs.\ \ref{Fig3}(g-l)], and plasmonic-regime doping (i.e., with five additional electrons, $Q=5$).

Independent of doping, the absorption cross section of metallic chains remains relatively unchanged by increasing the optical intensity, an observation compatible with their consistently smaller nonlinear polarizabilities. In contrast, the effective polarizabilities of the insulating phases offer a larger nonlinear response, where in particular Figs.\ \ref{Fig3}(c), (h), and (n) present strong saturation and also shifting of the peak energy, which eventually should transition towards a bistable regime. This behavior is corroborated by the large corresponding Kerr polarizabilities [Figs.\ \ref{Fig3}(f), (k) and (q), respectively], with the real part determining the peak shift strength and direction (e.g., red- vs blue-shift), while the imaginary part governs its saturation. Within the range of parameters considered here, we conclude that the TI is the most nonlinear material without doping, presumably because of the localized spatial and spectral character of the edge states (i.e., intrinsic anharmonicity), while the population of its edge states and subsequent Pauli blocking through doping renders its nonlinearity comparable to that of the insulator. The strong nonlinearity of the two types of doped insulators compared with the metallic chain can be understood in terms of a sparse Fermi sea, where interactions are poorly screened.

\begin{figure*} 
	\begin{center}
		\includegraphics[width=1\textwidth]{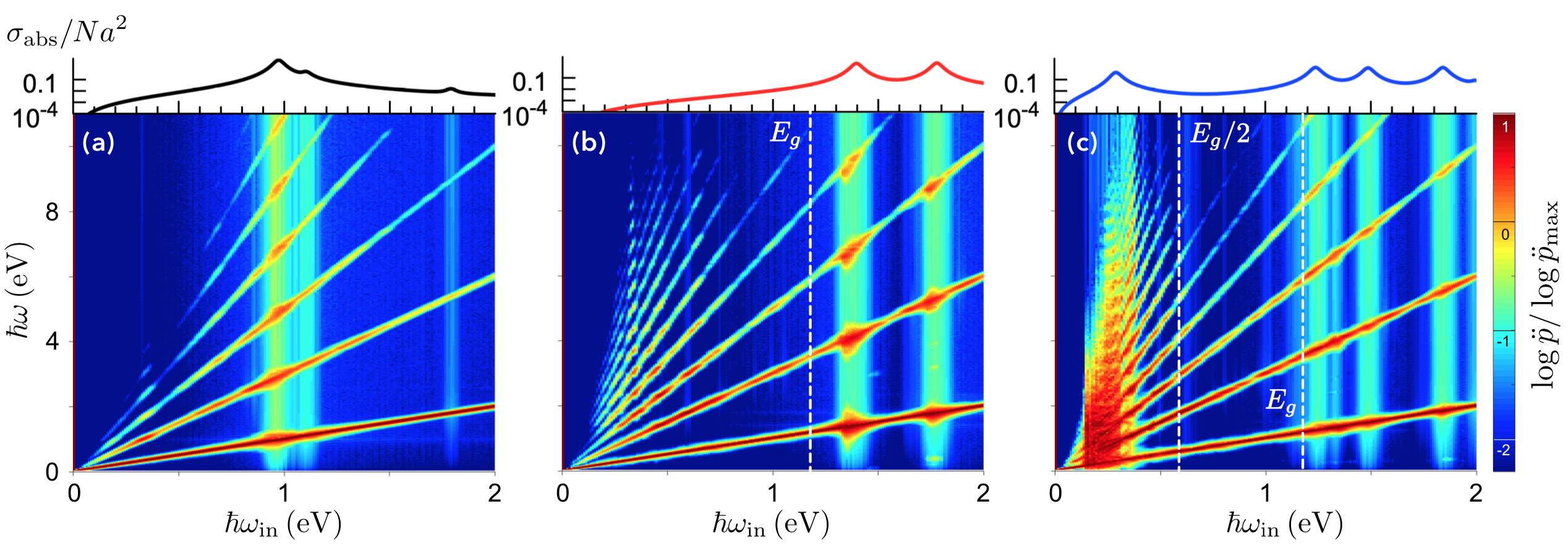}
		\caption{{\bf High harmonic generation in undoped SSH chains.} {\bf (a-c)} Induced dipole acceleration $\da$ in undoped SSH chains as a function of the output frequency $\omega$, as generated by a Gaussian pulse of peak intensity $10^{13}$\,W/m$^2$, 100\,fs FWHM duration, and central frequency $\omega_{\rm in}$ for the (a) metal, (b) insulator, and (c) topological insulator 50-atom SSH chains. The inset above each panel shows the respective linear response of the chain, while the vertical dashed lines mark the band gap energy $E_g$ and its half-value $E_g/2$.
		} 
		\label{Fig4}
	\end{center}
\end{figure*}

\section{High-harmonic generation in SSH chains}

\begin{figure*} 
	\begin{center}
		\includegraphics[width=1\textwidth]{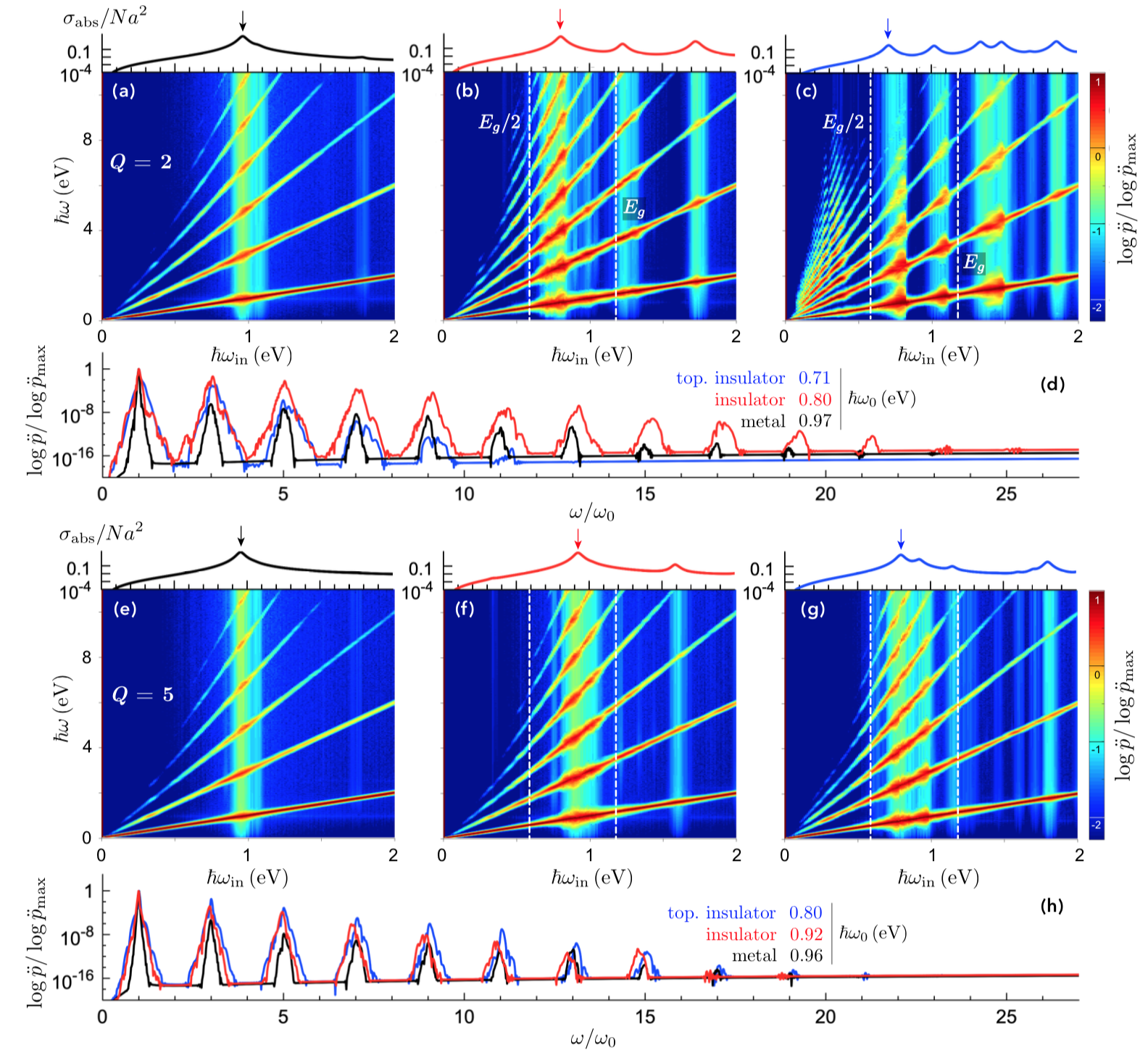}
		\caption{{\bf High harmonic generation in doped SSH chains.} Same as Fig.\ \ref{Fig4} when the chain is doped with either 2 (a-c) or 5 (e-g) electrons. Panels (d) and (h) show the normalized induced dipole at the resonances pointed by the arrows in the linear response for the metal (black), the insulator (red), and the topological insulator (blue) in (a-c) and (e-g), respectively.
		} 
		\label{Fig5}
	\end{center}
\end{figure*}

We further study the optical response of SSH chains to strong ultrashort laser pulses. In Figs.\,\ref{Fig4} and \ref{Fig5} we plot the induced dipole acceleration $\da$ normalized to the maximum dipole acceleration $\da_{\rm max}$ for each input frequency $\omega_{\rm in}$ considering Gaussian pulses of 100\,fs FWHM duration and peak intensity $I=10^{13}$\,W/m$^2$, with the horizontal pulse carrier frequency given on the horizontal axes and the frequency component of $\da$ indicated on the vertical axes; each contour plot of $\da$ is supplemented by the associated linear optical response of the system under consideration (upper panels). In the insulating phases we use white vertical dashed lines to indicate the electronic band gap energy $E_{\rm g}$ and $E_{\rm g}/2$, with the latter quantity indicating the edge-state-to-LUMO transition energy for the TI.

In Fig.\ \ref{Fig4} we consider undoped SSH chains, which present prominent features along the $\omega=s\omega_{\rm in}$ curves, where $s$ is an odd integer (i.e., harmonic generation for $s>1$). In all cases we observe that strong HHG is produced where the optical resonances intersect with the impinging light frequency, particularly in the low-energy feature observed in the absorption spectrum of the TI (see Fig.\ \ref{Fig4}c) lying below its band gap and associated with its topologically protected states, for which more high-order harmonics can couple to interband transitions.

In Fig.\ \ref{Fig5} we study HHG in LUMO-doped [Fig.\ \ref{Fig5}(a-d)] and plasmonic-regime-doped [Fig.\ \ref{Fig2}(e-h)] chains. The general trends for HHG are similar to those observed for Kerr polarizabilities in Fig. \ \ref{Fig3} and discussed in the previous section. As in the case of monochromatic excitation, we once again observe that the metal nonlinear response remains relatively unaffected by doping, while the insulator and the TI present higher HHG yields. In panels (d) and (h) of Fig.\ \ref{Fig5} we compare $\da$ at the dominant optical resonance of each SSH [a frequency denoted $\omega_0$, see arrows in the upper panels of Figs.\ \ref{Fig5}(a-c) and (e-g)]. For $Q=2$ [Fig.\ \ref{Fig5}(d)] the insulator presents the best harmonic yield, producing sizable peaks up to $\sim25\omega_0$, with harmonics exhibiting a slight blueshift associated with the pulse self-interaction involving the out-of-equilibrium electrons that it excites.\cite{paper305} In contrast, at $Q=5$ doping the insulator and TI produce very similar HHG yields that extend up to $\sim21\omega_0$, with the harmonics produced by the insulator slightly redshifted for $\omega>9\omega_0$. Both types of insulators display a harmonic yield much stronger than the metal. As already discussed in the previous section, the strong HHG of the doped insulators can be ascribed to weak screening in a low-density electron gas.

\section{Beyond the SSH model: High harmonic generation in carbon nanotubes}

To validate the predictive capabilities of the SSH model, we turn now to a more realistic 1D system in which to study HHG. Carbon nanotubes (CNTs) constitute a material platform that can behave as a metal, insulator, or TI, depending on chirality, and furthermore, like graphene, their electronic properties can be reasonably well-described by a tight-binding Hamiltonian. CNTs themselves hold great potential for diverse applications\cite{VTB15} because of their excellent mechanical, electronic, and optical properties, exemplified through the recent demonstration of a functioning CNT-based transistor.\cite{HLW19} In the field of nano-optics, recent experimental studies have explored the low-energy plasmons supported by CNTs when they are electrically doped,\cite{HFT18,FCF17,SAH15} similar to the collective excitations in highly doped graphene\cite{FAB11}, motivating their application for nanophotonic devices and nonlinear plasmonic elements.\cite{paper337} From a theoretical perspective, plasmons in CNTs have been extensively studied using both {\it ab initio} methods\cite{SAS99,DNA11} and also semi-classical approaches\cite{AFP08,paper264,paper278} based on the RPA\cite{PB1952} to calculate their optical conductivities. Incidentally, it has been experimentally proven that electrons in CNTs behave as Luttinger liquids,\cite{BCL99,IKS03} which have special relevance at low temperatures and are qualitatively corroborated by our methods.

\begin{figure} 
	\begin{center}
		\includegraphics[width=0.47\textwidth]{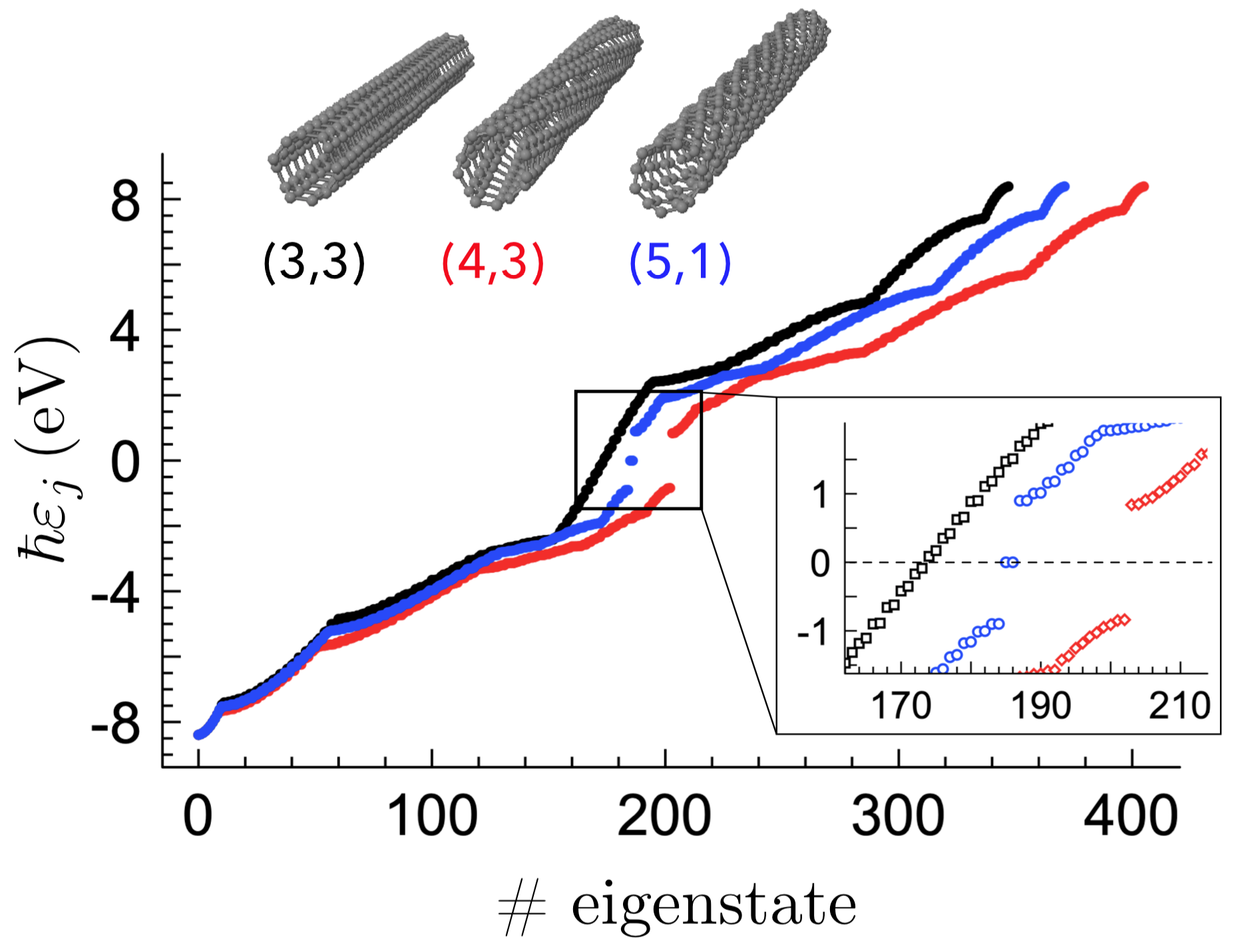}
		\caption{{\bf Band structure of CNTs.} {Single-electron energies for CNTs with chiralities (3,3), (4,3), and (5,1), corresponding to a metal (black squares), insulator (red diamonds), and topological insulator (blue circles), respectively. The inset shows the region near zero energy (indicated by the dashed horizontal line) in more detail.}} 
		\label{Fig6}
	\end{center}
\end{figure}

CNTs are constructed by wrapping a graphene layer into a cylindrical surface; a carbon atom at the origin is then identified with one at the graphene lattice position $n\ab_1 + m\ab_2$, where $\ab_1$ and $\ab_2$ are the conventional graphene lattice vectors, while the pair of integers $(n,m)$ determine the chirality of the tube. The resulting CNT diameter (in nm) is\cite{OHK00} $d\approx78.3\times10^{-3}\sqrt{ n^2 + m^2 + nm}$. We can additionally classify CNTs depending on their topology, which incidentally is determined by $n$ and $m$: when $n=m$ the CNT is metallic, if $n=m+1$ we have an insulator, and otherwise the CNT is a TI.\cite{OIE19,LKS15} To explore 1D-like structures more similar to SSH chains, we choose extremely thin CNTs experimentally reported with different chiralities.\cite{T17,HKM03,GSI08} Also, to facilitate the comparison with our 50-atom SSH chains, we take them to be roughly 7\,nm long. In particular, we consider CNTs with chiralities (3,3), (4,3), and (5,1), which have diameters of 0.41\,nm, 0.48\,nm, and 0.44\,nm, and correspond to a metal, an insulator, and a topological insulator, respectively. In Fig.\ \ref{Fig6} we show the band structures of these three CNTs calculated through a tight-binding model with a phenomenological nearest-neighbors hopping of 2.8\,eV,\cite{OHK00,CGP09} revealing $E_{\rm g}=1.80$\,eV for the TI and $E_{\rm g}=1.68$\,eV for the insulator.
 
\begin{figure*} 
	\begin{center}
		\includegraphics[width=1\textwidth]{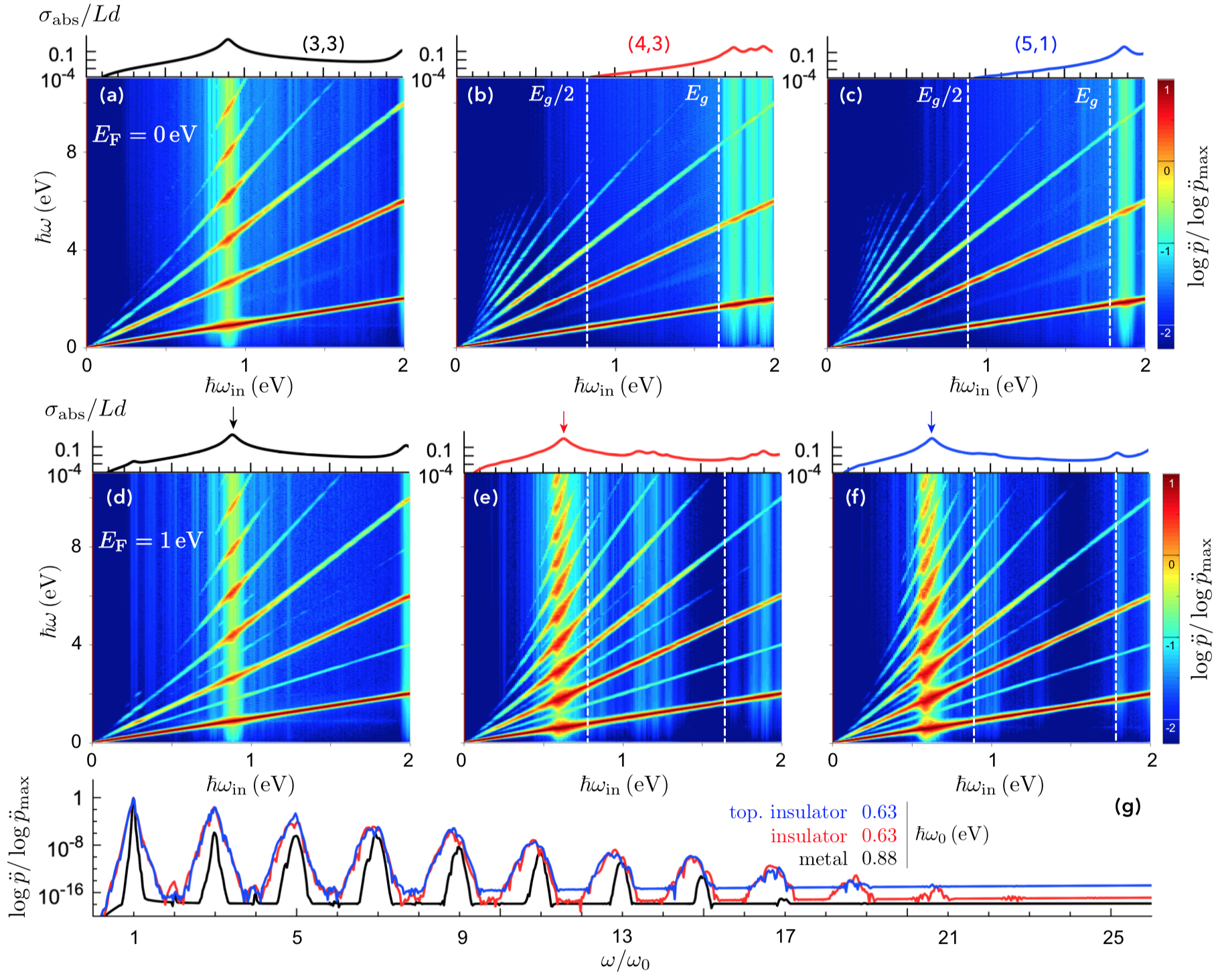}
		\caption{{\bf High harmonic generation in CNTs} {\bf (a-c)} Dipole acceleration $\da $ as a function of the output frequency $\omega$ generated by a Gaussian pulse of peak intensity $I=10^{13}$\,W/m$^2$, 100\,fs FWHM duration, and central frequency $\omega_{\rm in}$. The insets above each panel indicate the associated linear response for the (a) metallic, (b) insulating, and (c) topological insulating undoped CNTs. We have marked with white vertical dotted lines both the energy of the band gap $E_g$ and half the energy of the band gap $E_g/2$. {\bf (d-f)} Same as (a-c) but for 1\,eV-doped CNTs. The arrows indicate the resonant input frequency $\omega_0$ of the Gaussian pulse at which there is a boost in HHG.  {\bf (g)} Normalized dipole acceleration at the resonances pointed by arrows in the linear response on top of panels (d-f).}
		\label{Fig7}
	\end{center}
\end{figure*}

Strong-field driven electron dynamics in CNTs is simulated once again by inserting a tight-binding Hamiltonian into the equation of motion \eqref{eq:eom}, adopting a phenomenological damping rate of $\gamma=50$\,meV, and applying 100\,fs FWHM laser pulses of intensity $I=10^{13}$\,W/m$^2$ to excite high harmonics in these structures. In Fig.\ \ref{Fig7} we compare the HHG yields of undoped [Fig.\ \ref{Fig7}(a-c)] and doped ($\Ef=1$\,eV) CNTs [Fig.\ \ref{Fig7}(d-g)], again supplementing contour plots of $\da$ with linear absorption spectra (upper panels). For undoped CNTs we observe weaker HHG for both the insulator and the TI, only exciting up to the 7$^{\rm th}$ order, in contrast to the metallic tube, exhibiting a distinctly higher HHG yield, particularly when the impinging light energy coincides with the dominant optical resonance near 0.9\,eV. This particular trend is not followed by SSH chains, which in the undoped configuration produce stronger yield in insulators compared with metals [cf. Figs.\ \ref{Fig4}(a-c) and Figs.\ \ref{Fig7}(a-c)].

The HHG yield is strongly enhanced by doping the CNTs to a Fermi energy $\Ef=1$\,eV [Fig.\ \ref{Fig7}(d-g)], which introduces localized plasmon resonances. By inspecting the high harmonics generated at resonant frequencies when excited at optical resonances [Fig.\ \ref{Fig7}(g)] we see that both the insulator and the TI have higher yields, that their harmonics are quite strongly redshifted beyond the 7$^{\rm th}$ order, and that their cutoff is at the 21$^{\rm st}$ harmonic; this behavior was qualitative predicted by the simpler SSH model.

\section{Conclusions}

Despite the impressive pace at which the field of solid-state HHG is developing in both experiment and theory, the ideal material platform in which to realize this extreme nonlinear optical process has yet to be identified. Optical resonances supported by materials with intrinsically different types of electronic structure constitute an underexplored possibility to enhance the electric fields driving HHG, which we address here through the use of an intuitive model that contains much of the relevant physics. More precisely, our main conclusions based on the model SSH 1D chain are corroborated in their more realistic carbon-based analogues. In metals or doped semiconductors, where free electrons are present, self-consistent electron interactions become extremely important in both the linear and nonlinear response, and not only when dealing with optical resonances. This effect is stronger for doped semiconductors than for metals because weaker screening in the former makes their interactions more important. While HHG appears to be most efficient in semiconductors for harmonics generated below the band gap, the addition of a small amount of doping charge can produce a poorly screened, low-density electron gas with an intraband plasmon excitation that falls in this regime, therefore concentrating the impinging electromagnetic fields and {further boosting the HHG efficiency. This finding suggests the exploration of highly doped semiconducting materials as a promising platform for solid-state HHG. Our results pave the way for further investigation on the effects of electron-electron interactions in solid-state HHG, elucidating the involved microscopic mechanism and the relation between electronic band structure and the HHG yield, thus supporting its application towards nonlinear plasmonics, topological optoelectronics, and all-optical time-resolved probing of topological phases.

\acknowledgements
This work has been supported in part by the Spanish MINECO (MAT2017-88492-R and SEV2015- 0522, and FIS2017-84368-P), the ERC (Advanced Grant 789104-eNANO), the European Commission (Graphene Flagship 696656), the Catalan CERCA Program, and the Fundació Privada Cellex, and Universidad Complutense de Madrid (grant No. 962085). S.\ d.\ V.\ acknowledges financial support through the FPU program from the Spanish MECD. The Center for Nano Optics is financially supported by the University of Southern Denmark (SDU 2020 funding). J.\ D.\ C. was supported by VILLUM Fonden (grant No. 16498).


\end{document}